# L'Astronomia del Venerdi' Santo e l'ora della Sindone

Costantino Sigismondi

Ateneo Pontificio Regina Apostolorum

a mia madre, Erminia, che è nata 3 giorni prima della fine

dell'Anno Santo della Redenzione del '33, alle 3 di notte

**Abstract**

The solar eclipse of Good Friday has always been recognized as a problem from the astronomical point of view. The date of Crucifixion has been investigated by many scholars and the most accredited one is friday 3 april 33, when, at sunset an eclipsed Moon rised. Few minutes later the apparition of a third star of medium magnitude stated the beginning of the sabbatical rest: before that time Jesus has been buried, wrapped into the shroud. All these topics are reviewed from an astronomical point of view with some theological considerations.

**Sommario**

L'eclissi del Venerdì Santo è sempre stata riconosciuta come un problema dal punto di vista astronomico. La data della Crocifissione è stata investigata da molti studiosi, e la più accreditata è il venerdì 3 aprile 33, quando al tramonto sorse una Luna in eclissi. Pochi minuti dopo l'apparizione della terza stella di media grandezza, sanciva l'inizio del riposo sabbatico. Prima di quell'istante Gesù sarebbe dovuto essere sepolto, avvolto nella Sinodne. Tutti questi argomenti sono rivisti alla luce dell'astonomia.

# 1. Introduzione

Nell'anno liturgico C (come il 2013), la proclamazione del Vangelo della Passione del Signore durante la Domenica delle Palme segue il testo di Luca. In questo testo viene dichiarato esplicitamente che il Sole si eclissò in concomitanza con la morte del Salvatore. Gli altri sinottici non menzionano l'eclissi, e neppure il quarto Vangelo.

La querelle astronomica sulla possibilità di avere un'eclissi di Sole durante la fase di Luna piena è sorta già nel II secolo, subito dopo la stesura dei Vangeli, con Giulio Africano. Vedremo come si puo' inquadrare e risolvere nel panorama storico-religioso e culturale del tempo, confrontandola pure con i testi delle catechesi di san Cirillo di Gerusalemme.

Anche l'ora in cui entra in scena la Sindone, il tempo della sepoltura di Gesu', è stata regolata dall'astronomia: con la comparsa nel cielo di una terza stella di media grandezza iniziava il riposo dello Shabbat. Gli ebrei, infatti, avevano la loro vita quotidiana, e soprattutto le feste e le vigilie, ritmate da fenomeni astronomici osservabili.

Nel secondo capitolo studiamo la data del Venerdì Santo, nel terzo l'eclissi di Luna verificatasi il 3 aprile del 33, che è la data più accreditata per la Crocifissione, nel quarto capitolo è la Luna nelle catechesi di Cirillo di Gerusalemme, nel quinto la questione dello «splendere delle luci del Sabato», del riposo festivo alla deposizione di Gesu' dalla Croce sulla Sindone.

Il sesto capitolo è dedicato alle catechesi di Cirillo sul Sole e in particolare sulla tenebra del Venerdì Santo, e il settimo all'eclissi nel Vangelo di Luca e alla sua spiegazione.

## 2. La data del Venerdì Santo

Nel 1933 in occasione dell'anno santo della Redenzione indetto il 2 aprile, ultima domenica di Quaresima, dal papa Pio XI i padri gesuiti della Specola Vaticana e dell'Istituto Biblico hanno svolto uno studio sulla cronologia di Cristo, al fine di stabilire la disputa attorno alle due possibili date per la passione di Gesù. Alla fine risolsero per l'anno 33, ed i loro risultati sono stati pubblicati in un libro in latino, la *Chronologia Vitae Christi*.[1] Queste conclusioni sono accolte dalla maggior parte degli storici e soddisfa alla condizione di essere un venerdì 14 Nisan.[2]

Le questioni sulla cronologia della vita di Cristo sono vaste[3] e qui scelgo, con l'astronomo gesuita Gustav Teres[4] (p. 213), di presentare la tabella cronologica con le date più probabili, compilata dal Corbishley.[5]

---

[1] U. Holzmeister, *Chronologia Vitae Christi, Chronologia Mortis Domini*, Pontificio Istituto Biblico (1933).

[2] Essendo una data mobile, legata alla Luna, che il primo del mese veniva avvistata nella sua fase crescente iniziale, non corrisponde ad un giorno fisso della settimana. La probabilità, riscontrabile su periodi lunghi di anni, che il 14 Nisan cada di Venerdì è 1/7, tuttavia la coincidenza capitò sia il 7 aprile del 30 che il 3 aprile del 33. Su queste due date si è concentrata l'attenzione degli studiosi per la determinazione del Venerdì Santo.

[3] La stessa cronologia interna ai Vangeli sulla durata della Passione apre varie discussioni. Cito come esempio una cronologia diversa: morte di Gesù il mercoledì 14 nissàn, resta nel sepolcro 3 giorni e 3 notti come Giona, 14-15,15-16 e 16-17 e risorge all'alba del 18. Mercoledì, 14 nissàn. Prima che faccia buio Gesù è posto nel sepolcro. Giovedì, 15 nissàn. "Sabato" (giorno festivo). Gesù è nel sepolcro: al tramonto si compie il primo giorno e la prima notte. Venerdì, 16 nissàn. "Passato il sabato, Maria Maddalena e Maria Giacomo e Salome comprarono aromi" (Mc 16,1). "Poi tornate, prepararono aromi e profumi". – Lc 23,56. Gesù è nel sepolcro: al tramonto si compie il secondo giorno e la seconda notte. Sabato, 17 nissàn. Sabato settimanale. "Il sabato si riposarono, secondo il comandamento". – Lc 23,56. Gesù è nel sepolcro: al tramonto si compie il terzo giorno e la terza notte. Gesù viene resuscitato. Domenica, 18 nissàn. Primo giorno della settimana (nostra domenica): le donne trovano la tomba vuota. L'ipotesi è suggestiva per il compimento dei 3 giorni nel sepolcro come Giona e per l'attività delle donne con gli aromi, ma manda a pallino sia gli studi esegetici che tutta la cronologia, poiché il mercoledì 14 nissàn arriva solo negli anni 34 o 35, a seconda dell'avvistamento della prima falce di Luna crescente. A queste datazioni giunse anche Sir Isaac Newton che studiò approfonditamente l'argomento nel 1733 e l'algoritmo per ricostruire il calendario giudeo in quegli anni. Newton propendeva per l'anno 34. Newton, I., *Of the Times of the Birth and Passion of Christ*, chapter 11 in *Observations upon the Prophecies of Daniel and the Apocalypse of St. John* (London: J. Darby and T. Browne), pp. 144-168 (1733). Si veda anche l'articolo di John P. Pratt, *Newton's Date For The Crucifixion*, Quarterly Journal of Royal Astronomical Society **32**, 301-304 (1991). anche su http://www.johnpratt.com/items/docs/newton.html

[4] G. Teres, *The Bible and Astronomy, The Magi and the Star in the Gospel*, Springer Tudomanyos Kiado Kft. (1999³).

[5] T. Corbishley, *Chronology of New Testament Times*, in *The New Catholic Commentary on Holy Scripture*, Thomas Nelson & Sons Ltd, London (1975).

| Chronologia Vitae Christi ||
|---:|:---|
| a. C. 8/7 | Primo censimento in Giudea e nascita di Gesù |
| 6 | Fuga in Egitto e strage degli innocenti |
| 4/3 | Ritorno a Nazareth |
| d. C. 28 | in autunno, o primavera del 29, Giovanni il Battista |
| 29 | Battesimo di Gesù ed inizio del suo ministero pubblico |
| 30 | Prima Pasqua pubblica in Gerusalemme (7 aprile) |
| 31 | Seconda Pasqua in Gerusalemme (25 aprile) |
| 32 | Terza Pasqua in Gerusalemme (14 aprile) |
| 33 | Passione Morte e Risurrezione (3-5 aprile) |
| 34 | in autunno o primavera del 35 martirio di Stefano e conversione di Saulo |

Dunque la data più accreditata del Venerdì Santo è il 3 aprile del 33.

Questa stessa data è stata oggetto di studio da parte di due ricercatori inglesi Humpreys e Waddington[6] che nel 1983 hanno proposto di datare la crocifissione con un'eclissi parziale della Luna piena di Pasqua, accaduta venerdì 3 aprile 33 d. C. e visibile a Gerusalemme dal momento del sorgere della Luna, nella sua fase terminale quando solo il 20 % della superficie della Luna era ancora in eclissi. La tesi di questi studiosi è che questa eclissi fu vista da molti giudei che prima di consumare il pasto di Pasqua attendevano il sorgere della Luna (alle 18:20), già arrossata per effetti di estinzione della luce vicino all'orizzonte ed in più con un bordo superiore sanguigno dovuto alla fase finale dell'eclissi (terminata alle 18:50). A questo fenomeno, sempre secondo i due scienziati, si sarebbe riferito Pietro nel suo discorso di Pentecoste (At 2,20 = Gioele 3,4) "Il Sole si muterà in tenebra e la Luna in sangue" con conseguente grande consenso della folla che lo aveva notato poiché si trattava della Luna di Pasqua.

---

[6] J. Humpreys e W. G. Waddington, *Nature* **306**, 743-46 (1983).

## 3 L'eclissi di Luna del venerdì 3 aprile 33

Già nella mia tesi di Licenza in S. Teologia[7] avevo fatto notare che a mio avviso, se non si fosse trattato della Luna di Pasqua, questo fenomeno sarebbe certamente passato inosservato dai più, tanto più che proprio ad oriente di Gerusalemme (e ad Aprile Sole e Luna in opposizione al Sole sorgono esattamente ad Est) c'è il Monte degli Ulivi, di 818 metri sul livello del mare, che ostacola in parte la vista della Luna appena sorta, cioè proprio per la mezz'ora che restava ancora in eclissi; e che non era molto plausibile che tremila persone (quelli che si unirono agli Apostoli dopo il discorso di Pietro il giorno di Pentecoste) siano state sulla sommità del Monte degli Ulivi a osservare il sorgere della Luna prima di iniziare la cena pasquale.[8]

### 3.1 La visibilità della Luna al suo sorgere

Bradley E. Schaefer[9] afferma che la fase finale dell'eclissi, che occorreva proprio al sorgere della Luna quel venerdì 3 aprile 33, non si sarebbe potuta distinguere dal fondo cielo per due ragioni: la luminosità del fondo cielo stesso al tramonto del Sole e l'estinzione atmosferica verso l'orizzonte. Il lavoro è molto accurato, ed è anche supportato da un'osservazione fatta durante l'eclissi di Luna del 27 agosto 1988, durante la quale la Luna è stata persa di vista a 2.9° sopra l'orizzonte. Nel procedimento stima innanzitutto l'estinzione atmosferica per il sito di Gerusalemme pari a 0.28 magnitudini per massa d'aria. Poi valuta in 40 masse d'aria lo spessore di atmosfera da attraversare all'orizzonte. Sulla base di ciò afferma che la Luna al suo sorgere e fino a quasi 4° di altezza non sarebbe stata visibile, e che il colore rosso non sarebbe stato determinato dall'eclissi, ma dall'estinzione atmosferica, risultando uguale a quello di qualunque altra Luna (piena e non) al suo sorgere.

---

[7] C. Sigismondi, *Tematiche astronomiche nei Vangeli in prospettiva cristologica* [relatore prof. Romano Penna], Pontificia Università Lateranense, (1998). Di questa tesi riporto, per completezza, in questo lavoro alcuni paragrafi relativi alla questione dell'eclissi del Venerdì Santo. Le parti non modificate sono riprodotte in carattere e spaziatura più piccolo.

[8] I Samaritani andavano sul monte Garizim a celebrare la Pasqua, e da lì osservavano con precisione il tramonto del Sole. Cfr. M. Baillet, *La Pasqua ebraico-samaritana nel 1986*, Le Monde de la Bible ??? 27-32 (1986). Traduzione in italiano di P. Servi e commentata da G. Berbenni nelle dispense DSS001_06 del corso di diploma in studi sindonici. Va notato come nella dinamica del sacrificio degli agnelli pasquali l'osservazione della Luna, che è quasi piena, ed è quindi già sorta alle 15:57 al tramonto del Sole, non sia riportata dal Baillet. Il primo agnello viene sgozzato 2 minuti dopo il tramonto del Sole avvenuto alle 18:10. Gli orari dei fenomeni astronomici del 22 aprile 1986 sono stati ricontrollati da me con il software di effemeridi Ephemvga.

[9] Schaefer, B. E., *Lunar visibility and the crucifixion*, Q. Jl. R. astr. Soc. **31**, 53-67 (1990).

Senza contestare il procedimento di Schaefer, devo però far notare che le 40 masse d'aria all'orizzonte possono essere esagerate per Gerusalemme. Per un errore, forse tipografico, Schaefer in quell'articolo pone Gerusalemme a 450 m sul livello del mare, e questo già comporta un airmass maggiore, ma in generale si possono avere condizioni molto migliori di trasparenza atmosferica, specialmente in una regione con poca umidità come la Giudea. Nei miei studi sulla trasparenza atmosferica a Ostia, sul mare, ho verificato come lo strato di umidità prossimo alla superficie, ed entro 20 metri di quota contribuisce come 60 masse d'aria, ma solo negli ultimi 15' sopra l'orizzonte, come mostra l'immagine del Sole al tramonto dell'11 agosto 2009, messa a confronto con l'alba a Gerusalemme dal Belvedere davanti al Muro del Pianto del 7 dicembre 2011.

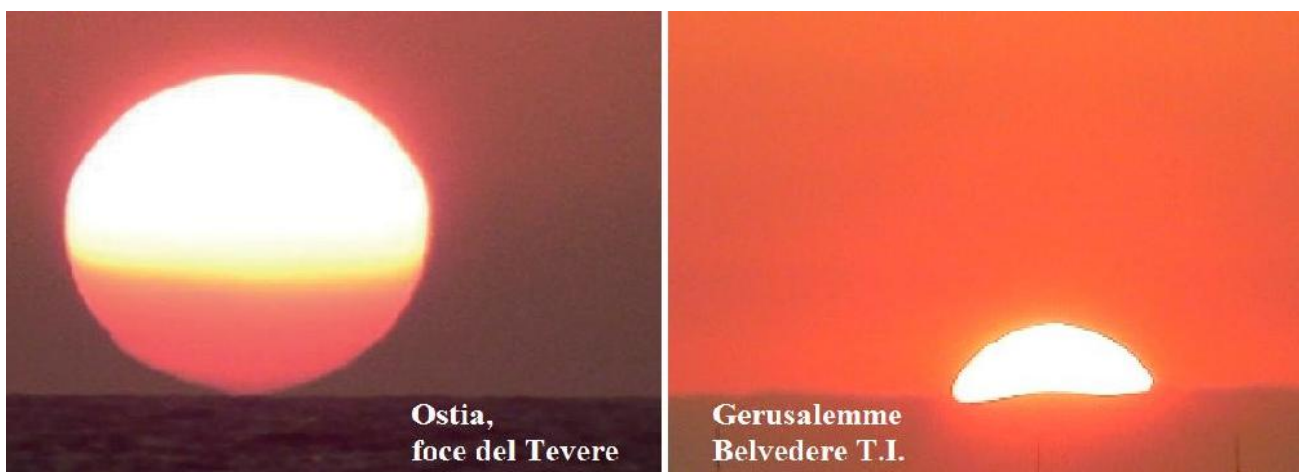

La foto ad Ostia è fatta con ISO 50 1/60s (SANYO HD1010) con lo zoom a 10x, quella a Gerusalemme ISO 50 1/53 s (SANYO CG9) lo zoom a 5x. E' evidente che l'estinzione atmosferica ad Ostia era molto maggiore di quella a Gerusalemme, nei primi 15' sopra l'orizzonte (16' sono equivalenti a metà diametro del Sole).

Per questa ragione, al di là della critica ai modelli numerici di estinzione atmosferica, ho programmato l'osservazione a Gerusalemme del sorgere della Luna piena del 28 novembre 2012, in eclissi penombrale e nei giorni immediatamente vicini, dal belvedere presso il Temple Institute di fronte al Muro del Pianto, dove ho già osservato lo spuntare del Sole sopra le alture del Negev, molto più in basso di Gerusalemme, dunque nella condizione di massimo assorbimento atmosferico, tra il 4 e il 10 dicembre 2011.

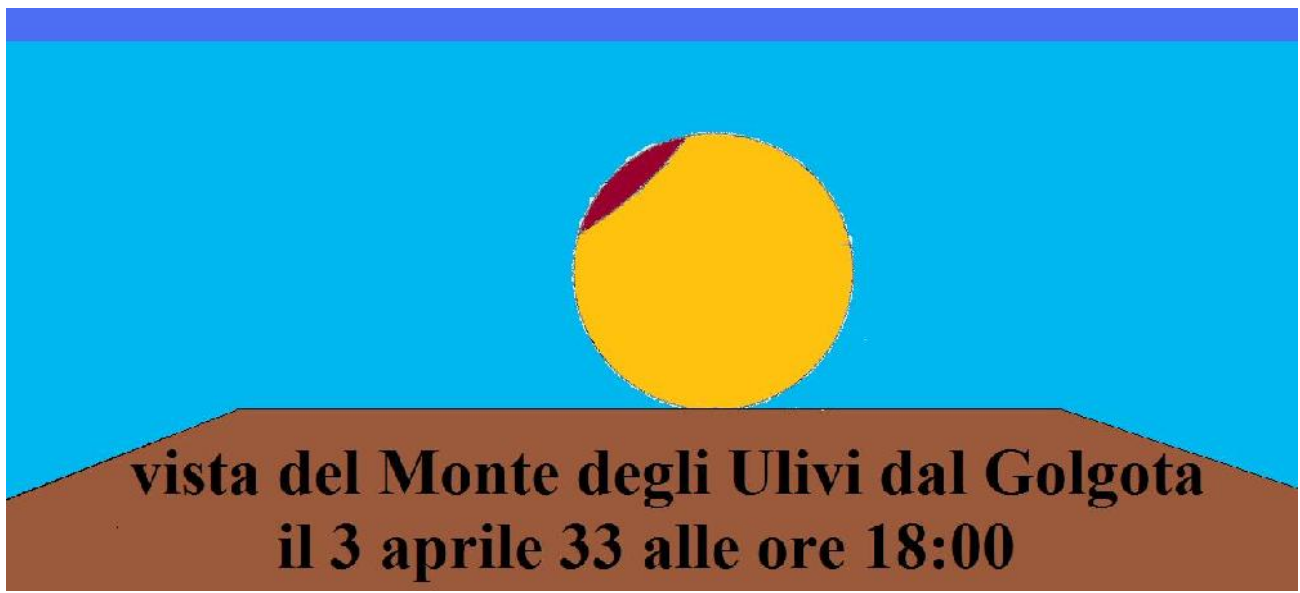

Tuttavia oggi mi ricredo su questa conclusione: l'altimetria della Città Santa poteva consentire di avvistare la Luna equinoziale abbastanza presto. Ad esempio dal Golgota a 779 m sul livello del mare, che si trova, in linea d'aria, a 1372 metri dalla verticale della sommità del Monte degli Ulivi (dove oggi si trova l'edicola dell'Ascensione)[10] si poteva osservare la Luna intera, inclusa la parte in eclissi, quando il suo bordo inferiore ha raggiunto l'altezza di 1°38'=arctan[(818-779)/1372], cioè alle 17:59, circa 10 minuti dopo il suo sorgere completo, la sera del 3 aprile 33.[11]

---

[10] Mi riferisco alle misure prese da satellite mediante google maps, riportate nella mia tesina per l'esame DSS003 del 15.11.2012. C. Sigismondi, *Il Santo Sepolcro orientamento della Basilica e le omelie di san Cirillo di Gerusalemme.*

[11] Con il software Ephemvga ho ricostruito gli orari del tramonto del Sole e del sorgere della Luna visti da Gerusalemme (32 N, 35 E) il 3 aprile del 33. La Luna sorge completamente sull'orizzonte Est alle 17:49, ma ancora non è visibile da Gerusalemme per via del Monte degli Ulivi che sta davanti alla Città Santa verso Est.
L'ultimo lembo di Sole scompare all'orizzonte alle 17:56. La condizione per avere un'eclissi lunare, è che Luna e Sole siano in opposizione (180°) in questo caso i centri dei due astri erano a 179.1° e l'eclissi già nella sua fase parziale terminale. Il diametro del cono d'ombra della Terra alla distanza della Luna è di circa 2°. La concomitanza di Luna e Sole in opposizione sopra l'orizzonte era stata già notata nell'antichità da Tolomeo (II sec. d. C.) che l'ha correttamente attribuita all'azione della rifrazione della nostra atmosfera che «solleva» gli astri fino a circa mezzo grado quando si approssimano all'orizzonte.

Dalla spianata del Tempio, che è più vicina al Monte degli Ulivi, solo 811 metri, l'altezza che il bordo inferiore della Luna doveva raggiungere era di 2°45'=arctan[(818-779)/811], quasi 16 minuti dopo che era sorta completamente, alle 18:05 in tempo per essere vista ancora in fase parziale di eclissi, come macchiata di sangue.

Rivedendo i calcoli dell'eclissi con il programma Occult 4, usato correntemente per le effemeridi di tali fenomeni astronomici si trova che la Luna quella sera uscì dall'ombra della Terra alle ore 18:11. La Luna rimaneva comunque nella penombra, cioè da tutta la sua superficie il Sole non era visibile completamente: questo ha conferito alla Luna un colore rossastro, dato che al termine dell'eclissi la Luna si trovava ad appena 4° sopra l'orizzonte orientale, quindi ancora sottoposta ad un certo assorbimento atmosferico. La fase di penombra, a causa della forte luminosità della Luna può passare anche inosservata,[12] poiché il confine tra penombra e illuminazione piena è continuo, a differenza dell'ombra.

Inoltre affermano ancora i due inglesi che la tenebra durata tre ore nel pomeriggio del Venerdì Santo, sarebbe stata provocata da una grossa tempesta di sabbia, che lasciò pulviscolo nell'atmosfera tanto da rendere particolarmente suggestivo il colore della Luna al suo sorgere. Su questa ultima considerazione sarei più cauto: il colore di una eclissi di Luna, chiamato anche indice di Danjon, è certamente dovuto alle condizioni dell'atmosfera terrestre, ma su larga scala, unitamente all'attività solare su scala undecennale.[13]

---

[12] Oltre alla visione ad occhio nudo ho usato due polarizzatori per attenuare la luce lunare ed ho visto con chiarezza la penombra dell'eclissi del 31 dicembre 2009. Paul Couderc nel libro *Les éclipses*, Presse Universitaire de France, Paris (1971), a p. 24 suggerisce di utilizzare un filtro per percepire l'avanzamento o l'arretramento della penombra sulla superficie della Luna. Couderc propone anche di osservare ad occhio nudo la Luna riflessa da due vetri che formano tra loro un angolo molto acuto: le riflessioni multiple ridurranno l'*éclat* della Luna in modo da percepire la penombra.

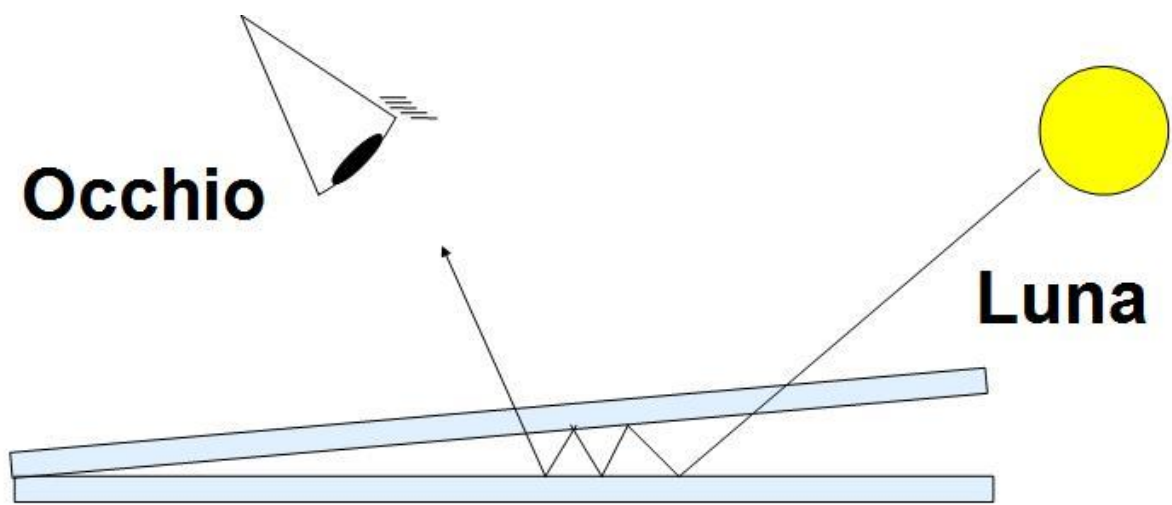

Nella figura successiva mostro dove nell'ombra della Terra è proiettata la zona dove si trova Gerusalemme. Per questo la correlazione tra intensità della "macchia di sangue" prodotta dall'ombra sulla Luna e la tempesta di sabbia non è pensabile, mentre quella tra passata tempesta e pulviscolo ancora in aria è più probabile, in tal caso è tutta la colorazione della Luna ad essere scurita.

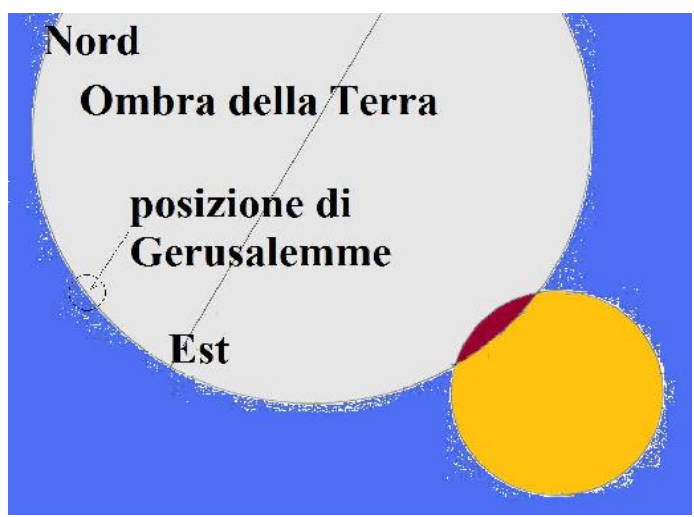

Nell'ultima parte dell'eclissi la Luna esce dal cono d'ombra della Terra, è dunque una parte molto piccola della Terra ad essere proiettata sul disco lunare, e non è quella corrispondente a Gerusalemme.

Concludendo: l'arrossamento della Luna puo' essere ascritto al pulviscolo atmosferico locale, ma non alla proiezione di questo stato dell'atmosfera terrestre sul bordo del cono d'ombra della Terra, poiché la Luna non lo intercetta dalla parte dove si trova Gerusalemme. L'eclissi contribuisce alla penombra, scurendo fino a 2 magnitudini lo splendore ordinario della Luna piena.

---

[13] P. Couderc, *Les éclipses*, Presse Universitaire de France, Paris (1971), p. 97-100.

Riporto la fotometria di un'eclissi di Luna nella figura seguente.[14] Si vede che la zona di penombra va da 2 magnitudini fino a zero man mano che ci si allontana dal centro dell'ombra.

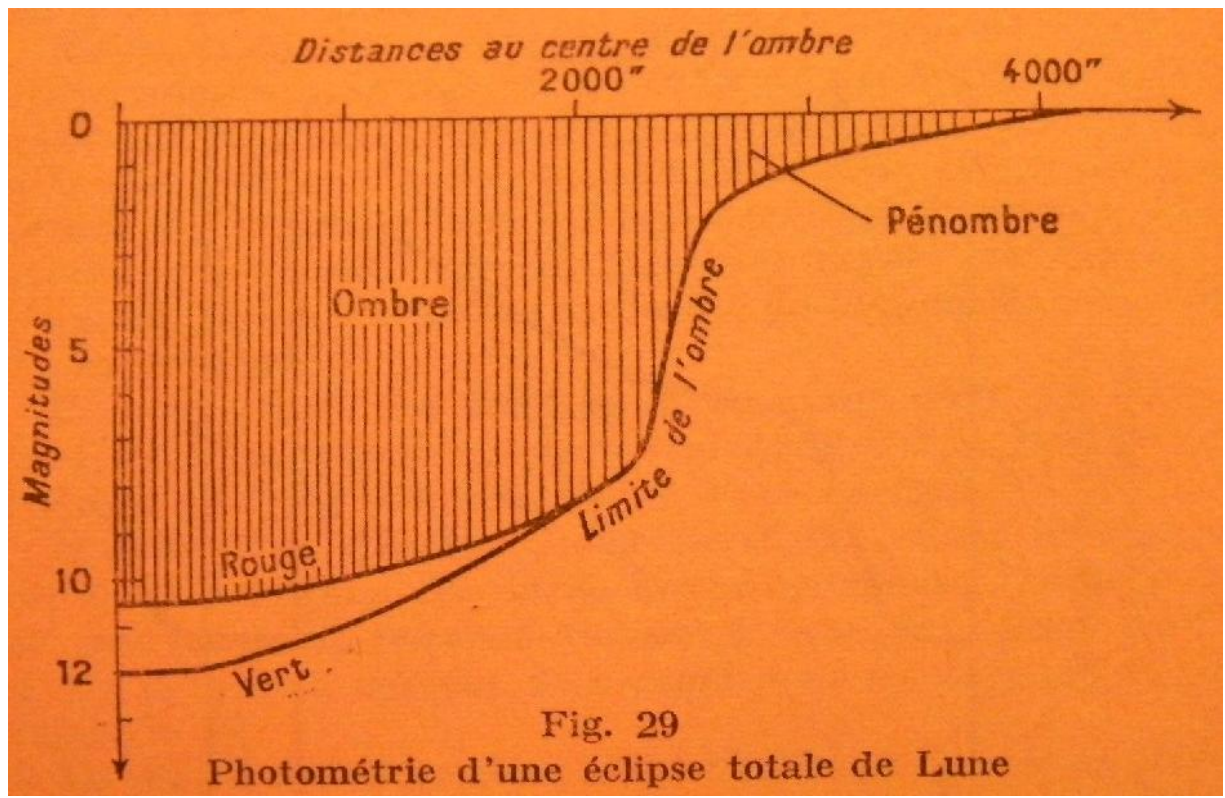

In ordinate lo splendore in magnitudini, la scala è diretta verso il basso per ricordare che una sorgente luminosa di prima grandezza (o magnitudine) è più luminosa di una di seconda e così via. Le magnitudini crescenti coripondono a delle luminosità più deboli. La magnitudine 10 corrisponde ad uno splendore 10000 volte più piccolo che la magnitudine 0. In ascisse le distanze dal centro dell'ombra in secondi d'arco: 1°=3600″.

La prossima eclissi lunare di penombra è il 28 Novembre 2012: la Luna esce dalla penombra alle 16:53 UT, ovvero alle 17:53 in Italia e alle 18:53 a Gerusalemme, dove sorge alle 16:29, quattro minuti prima del massimo dell'eclissi con 0.936 digits di magnitudo, ovverosia il 93.6% del diametro della Luna in penombra. L'azimuth è 65°, cioè 33° più a Nord del 3 aprile del 33. Proverò ad osservare vicino alla Menorah del Temple Institute, sopra il Muro del Pianto, così da avere ugualmente il Monte

---

[14] è la figura 29 di P. Couderc ( *op. cit.* 1971, p. 100) introvabile sui libri moderni, dove questi argomenti non vengono più trattati in profondità, e completa a meraviglia questa ricerca approfondita sull'eclissi lunare del 3 aprile 33.

degli Ulivi come sfondo.

Si potrebbero infine fare delle considerazioni di tipo statistico, sulla rarità di un fenomeno del genere, tuttavia esulano dal nostro contesto, in quanto delle date individuate per la crocifissione solo quella del 3 aprile 33 ha l'eclissi.[15]

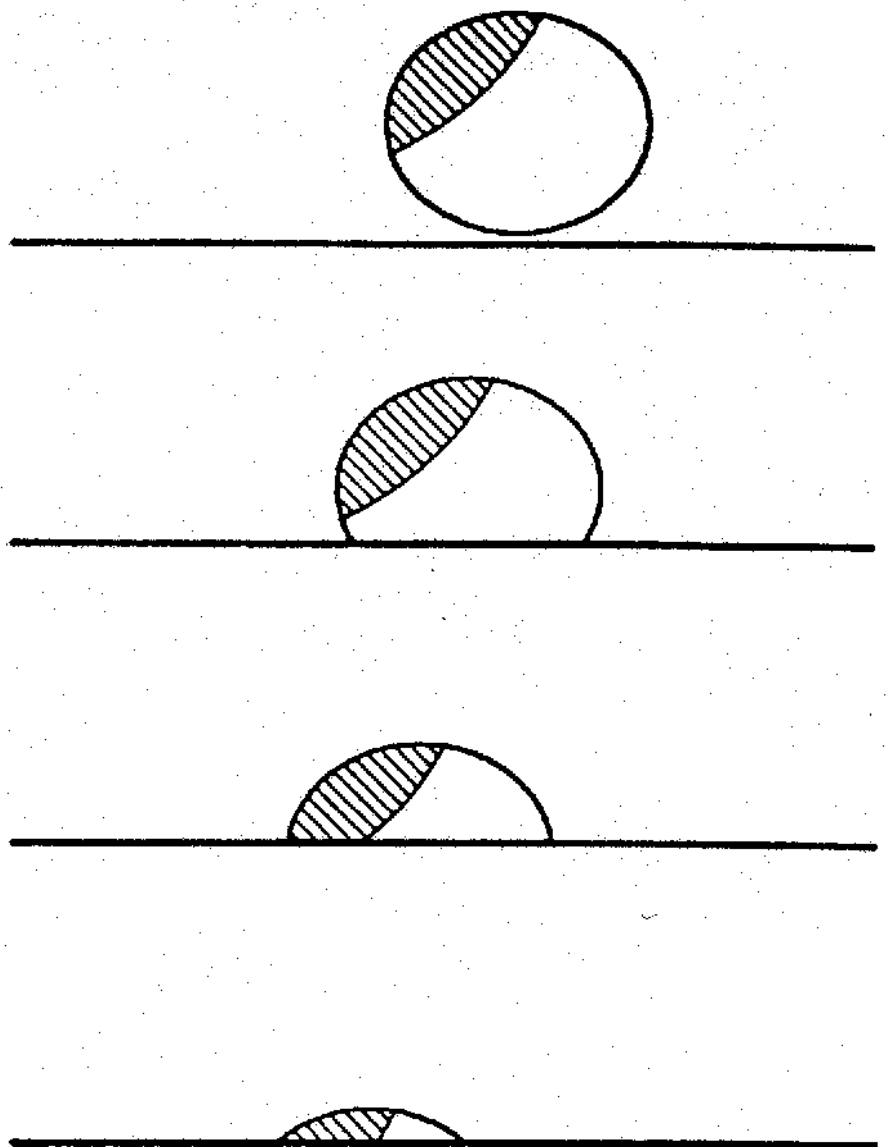

Il sorgere della Luna Venerdì 3 aprile del 33 d. C. come visto sull'orizzonte orientale (libero) di Gerusalemme. Gli effetti della rifrazione atmosferica sono stati inclusi e danno origine al profilo distorto della Luna. L'intervallo di tempo tra due successivi diagrammi (dal basso verso l'alto) è di 45 secondi. Il colore più probabile per la Luna era: rosso per l'area ombreggiata e giallo-arancio per l'area non ombreggiata (dall'articolo di J. Humpreys e W. G. Waddington, *Nature* **306**, 743-46 (1983).

---

[15] Altre considerazioni statistiche sulle eclissi si possono trovare in P. Zanna e C. Sigismondi, *Dùngal, letterato e astronomo. Note di stilistica e di astronomia sulla Lettera a Carlo Magno circa le eclissi di Sole dell'810,* Archivum Bobiense **XXVI**, 187-295 (2004). anche su www.arxiv.org/12113687.pdf

# 4. La Luna nelle catechesi di san Cirillo di Gerusalemme

È interessante vedere come il vescovo di Gerusalemme[16] recepiva le nozioni di astronomia riguardo alla Luna: questo traspare dalle sue catechesi battesimali.

Il Cristo dunque fu crocifisso per noi: giudicato quando era notte e il buio era a malapena rischiarato dai carboni che accesero per il freddo (cf. Gv 18, 18). Dopo che fu messo in croce all'ora terza, dall'ora sesta alla nona poi si fece buio (cf. Mc. 15, 25). All'ora nona sorse di nuovo la luce come predetto. Ma era stato profetato anche questo? Vediamolo.

L'aveva predetto il profeta Zaccaria: In quel giorno non vi sarà luce, sarà un giorno di freddo e di gelo – Pietro dovette difendersi dal gelo andando a riscaldarsi! –, un giorno noto al Signore (cf. Zac. 14, 6-7) – non al profeta che ne conosceva altri, non però questo della passione del Signore, giorno che fece il Signore (cf. Sal. 118 / 117, 24) –; un giorno noto solo al Signore, in cui non vi sarà né giorno né notte (cf. Zac. 14, 7).

Quale enigma ci propone il profeta parlando di un giorno che non è né giorno né notte? Come chiamare questo spazio di tempo?

La spiegazione è nel Vangelo che narra come si svolsero i fatti. Non fu un giorno come tutti gli altri, perché allora il sole non seguì il suo solito corso dall'Oriente all'Occidente, ma si oscurò e dall'ora sesta all'ora nona tutto fu scuro (cf. Mt. 27, 45), nel mezzo del giorno subentrò il buio che Dio chiamò notte: ecco perché dentro lo spazio d'un giorno non vi fu né giorno né notte, né una piena luce che potesse chiamarsi giorno né un totale buio che potesse chiamarsi notte (cf. Gen. 1, 5); ma il sole tornò a splendere dopo l'ora nona. Il profeta l'aveva predetto con le parole né giorno né notte alle quali aggiunse le altre ma alla sera vi sarà luce (cf. Zac. 14, 7). Vedi con quanta precisione i profeti preannunziarono quello che si sarebbe avverato! [Cat. XIII, 24]

Testimoni della Risurrezioni sono i tempi e i luoghi, angeli, uomini e cose, a cominciare dalla Luna piena.

Ma tanti altri testimoni ci segnalano la risurrezione del Signore. Partiamo da quella notte illuminata dalla luna piena, il sedici del mese, per venire alla roccia che gli offrì il sepolcro e a quella pietra che ivi lo sigillò alla presenza dei giudei. [Cat. XIV, 22]

In realtà il 16 del mese la Luna aveva già passato la fase di piena, essendoci stata l'eclissi che è la massima «pienezza» possibile per la Luna, quasi esattamente a 180° dal Sole. Dunque qui Cirillo non sta parlando con precisione scientifica, ma indica che la Luna è nella sua fase massima.

Ascolta le parole di Isaia che profetò: «Il cielo sarà aperto come un libro, e tutte le stelle cadranno come le foglie di una vite, come cadono le foglie da un fico», e quelle del Vangelo che recita: «Il sole si oscurerà, la luna perderà il suo splendore e gli astri cadranno dal cielo». Non affliggiamoci

---

[16] Riggi, C., *Cirillo di Gerusalemme, le Catechesi*, Città Nuova (1993) in corso di revisione da G. Berbenni.

quasi che dovessimo noi soli subire la morte, anche gli astri finiranno. [Cat. XV,4]

Questo è proprio il commento al Vangelo della XXXIII domenica del tempo ordinario.

Nel macrocosmo è la Luna figura della Risurrezione

Ma un'altra testimonianza della risurrezione dei morti puoi chiaramente prenderla tra i luminari del cielo.

Il corpo lunare infatti, dopo essere scomparso ai nostri occhi, al punto che non lo vediamo più neppure in parte, poi ridiventa pieno, precisamente come prima. Ma la luna offre una prova precisa di quel che stiamo dimostrando quando, dopo essere scomparsa sanguigna nella fase calante che capita periodicamente dopo un ciclo di anni, poi riprende il suo normale aspetto di corpo luminoso: questo fenomeno è stato disposto da Dio per te, perché anche tu uomo fatto di sangue non abbia a mettere in dubbio la risurrezione dei morti; quel che vedi avvenire nella luna si realizzerà anche in te. [Cat. XVIII, 10]

In questo passo Cirillo descrive gli effetti di due cicli noti come il Saros di 18 anni 10 giorni e 1/3 e quello di 18,6 anni della rotazione dei nodi dell'orbita lunare. Infatti la prima descrizione è proprio quella dell'eclissi totale. Il corpo lunare scompare e poi ridiventa pieno.

La fase calante di cui parla successivamente è meno chiara, visto che l'eclissi avviene a completamento della fase crescente. Poiché Cirillo aggiunge subito dopo che capita dopo un ciclo di anni, probabilmente allude proprio al ciclo che si completa in 18,6 anni. Agli estremi di questo ciclo ci sono i cosiddetti «lunistizi», ovversia delle posizioni estreme di sorgere e tramontare della Luna che vengono raggiunte solo ogni 18,6 anni. Anche nella nostra vita quotidiana del XXI secolo si possono riconoscere gli effetti di questo ciclo, ad esempio se cerchiamo i periodi in cui la Luna si alza di più sull'orizzonte. Per noi dell'emisfero Nord è la Luna di Natale la più alta dell'anno, ebbene da un anno all'altro si possono vedere le differenze: ci sono lune piene molto alte come nel 2008 e nel 2026 e ce ne sono di molto basse come come nel 1999 e nel 2017 e di intermedie in tutti gli altri anni. Un altro modo per vedere i lunistizi e notare i punti di tramonto o sorgere della Luna nel corso dell'anno.

Per l'uomo antico tutti questi fenomeni erano ben noti, a livello esperienziale, anche perché avevano meno ostacoli di noi ad osservare albe e tramonti.

La quasi coincidenza tra rotazione dei nodi e il Saros, periodo in capo al quale una eclissi si ripete, ha probabilmente suggertio a Cirillo di mostrare anche una «morte» e «risurrezione» della Luna su un periodo di tempo «di alcuni anni», cioè 18.

Vi è poi il ciclo di Metone, di 19 anni, usato nel computo della Pasqua cristiana, che pure avrà contribuito all'elaborazione di questo passo della diciottesima catechesi battesimale. Il ciclo di Metone afferma che la Luna piena avvenuta il 3 aprile 33 si ripete tal quale 19 anni dopo esatti, il 3 aprile 52.

Di seguito Cirillo ribadisce il valore di questi ragionamenti, in modo estremamente attuale:

Tu, utilizza questi argomenti con i pagani, perché con quelli che non credono nella Bibbia bisogna combattere con armi diverse da quelle che ci offrono le Scritture, soltanto ragionando e provando. Essi infatti ignorano Mosè e Isaia, non conoscono né il Vangelo né Paolo. [Cat. XVIII, 10]

Lo spirito del corso di diploma in studi sindonici, e del master in scienza e fede che si tengono alla Università Regina Apostolorum, da un punto di vista pastorale, soddisfa proprio questo consiglio del vescovo Cirillo.

Gli uomini e le donne dei nostri tempi, pagani di fatto o uomini di buona volontà, sembrano essere stuzzicati solo dai ragionamenti e dalle prove. In modo ancora più incisivo se si tratta di argomenti che riguardano l'astronomia. Il consiglio di Cirillo è quanto mai attuale.

C'è da dire però che, mutati i tempi, occorre anche mutare di argomentazioni. Le eclissi di Luna o di Sole non potrebbero più convincere sulla plausibilità della Risurrezione, che ci aspetta tutti, poiché in tutti questi eventi periodici non si intravvede l'unicità che invece riguarda l'uomo. Altrove Cirillo dice che gli alberi sperimentano la rinascita primaverile e perché non dovrebbero risorgere anche gli uomini, per i quali gli alberi sono stati creati? Interessante è poi è la citazione del mito greco della fenice, che ogni cinquecento anni si reca in una città d'Egitto per fare il suo nido, morirci e risorgere dalle sue putredini. Il mistero della vita e la sua comprensione per l'uomo antico è differente rispetto a noi: per loro era ammissibile la risurrezione «macroscopica» della fenice, poiché il concetto di generazione spontanea era forse implicito nella cultura del tempo, ma Cirillo riesce a fare un uso mirabile di questo mito, e con limpido ragionamento rafforza la speranza nella risurrezione finale per virtù di Colui che è autore dell'esistente dal nulla e della vita dall'inanimato.[17]

| 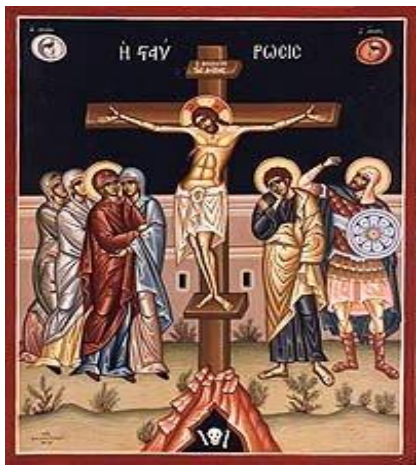 | In questa icona il Cristo apre le braccia tra l'oriente e l'occidente, tra la Luna che sorge ed il Sole che tramonta.<br><br>Sulla croce allargò le sue mani per abbracciare con il Golgota, posto proprio al centro della terra, tutto il mondo fino ai suoi estremi confini. Non sono io ad affermarlo, ma lo dice il profeta: Hai operato la salvezza dal centro della terra (Sal. 74 / 73, 12). [Cat. XIII, 28] |

---

[17] Cfr. C. Riggi, *Il mito dell'Araba Fenice in Cirillo di Gerusalemme*, in «Bessarione», 1989 (7), pp. 141-148.
Cirillo riporta: Ma i greci, in cerca ancora d'un segno chiaro della risurrezione dei morti, obiettano che questi esseri risorgono, non però dopo la loro totale putrefazione; vogliono constatare qualche chiaro caso di vivente che dopo la decomposizione sia risuscitato. Perciò Dio, cui è ben nota l'incredulità degli uomini, tra gli uccelli ne creò anche uno che ha nome fenice, di cui scrisse Clemente e molti novellarono come di un unico della specie, che ogni cinquecento anni vola verso le contrade d'Egitto e giuntovi fa vedere cosa sia risuscitare. Lo fa non in luoghi deserti, perché non vuole che il suo mistero rimanga nascosto, ma in una città ben nota perché tutti possano quasi toccare con mano l'evidenza incredibile del fatto.
Infatti al compiersi di detto numero di anni, la fenice entra nel nido che si è costruito con incenso mirra e altri aromi, e alla vista di tutti muore e marcisce. A questo punto dalla carne marcita del suo cadavere nasce una larva, che crescendo passa dalla forma di verme a quella di uccello. Il fatto non è incredibile, si verifica anche per le api che si formano da simili larve, anche per gli uccelli che sviluppano penne ossa e tendini a partire dagli elementi liquidi delle uova.
Così la fenice, messe le penne, torna infine a essere perfettamente quella di prima, e vola per l'aria come faceva prima di morire, dopo aver offerto agli uomini la più sicura testimonianza della risurrezione dei morti.
Uccello meraviglioso, la fenice! Ma essendo un uccello, animale irragionevole, non ha mai potuto cantare le lodi del Signore; vola alto per l'aria, ma senza raggiungere la conoscenza del Figlio di Dio, del vero Unigenito.
Dio dunque non darà la risurrezione a noi che ne cantiamo le lodi e ne osserviamo i comandamenti, dal momento che ha voluto che risorga dai morti un uccello privo di ragione, che non può riconoscere il suo creatore?
Ma poiché la testimonianza singolare della fenice ci viene da regioni lontane, da notizie peregrine alle quali per di più non si dà credito, torniamo a quelle tratte dal normale quotidiano che è sotto i nostri occhi.
Cento o duecento anni fa, dove eravamo tutti, noi che parliamo e voi che ascoltate? Non sappiamo come abbia avuto inizio la nostra natura corporea? Non sai che i princìpi della nostra generazione sono inerti, amorfi e indifferenziati? Eppure uno di tali princìpi indifferenziati e inerti prende forma e diventa un uomo vivo! Questo principio inerte diventa carne, e poi si trasforma in nervi vigorosi, in occhi luminosi, in naso sensibile agli odori, in orecchie capaci di ascoltare, in lingua che trascina con la parola, in un cuore che batte, in mani che lavorano, in piedi che agilmente camminano e in tutte le altre membra dalle forme più diverse. Questo principio inerte diventa capace di fabbricare navi, costruire case, elevare edifici, operare in ogni attività professionale, da militare o da magistrato, da legislatore e da re.
E Dio che ci ha creati a partire da elementi imperfetti non potrà risuscitare i morti? Chi ci ha formato il corpo a partire da quanto c'è di più vile, non potrà risuscitare questo corpo dopo che è morto? Colui il quale ha creato l'esistente dal nulla non potrà mediante la risurrezione restituirgli l'essere venuto meno? [Cat. XVIII, 9-10]

## 5. L'ora della Sindone: prima della terza stella

Per gli ebrei il riposo festivo è regolato dall'apparizione della terza stella[18] di media grandezza nel cielo serotino.

Le luci del Sabato sono dunque le stelle che ne testimoniano l'inizio, oppure i fuochi accesi prima del riposo di precetto?

Vediamo con ordine.

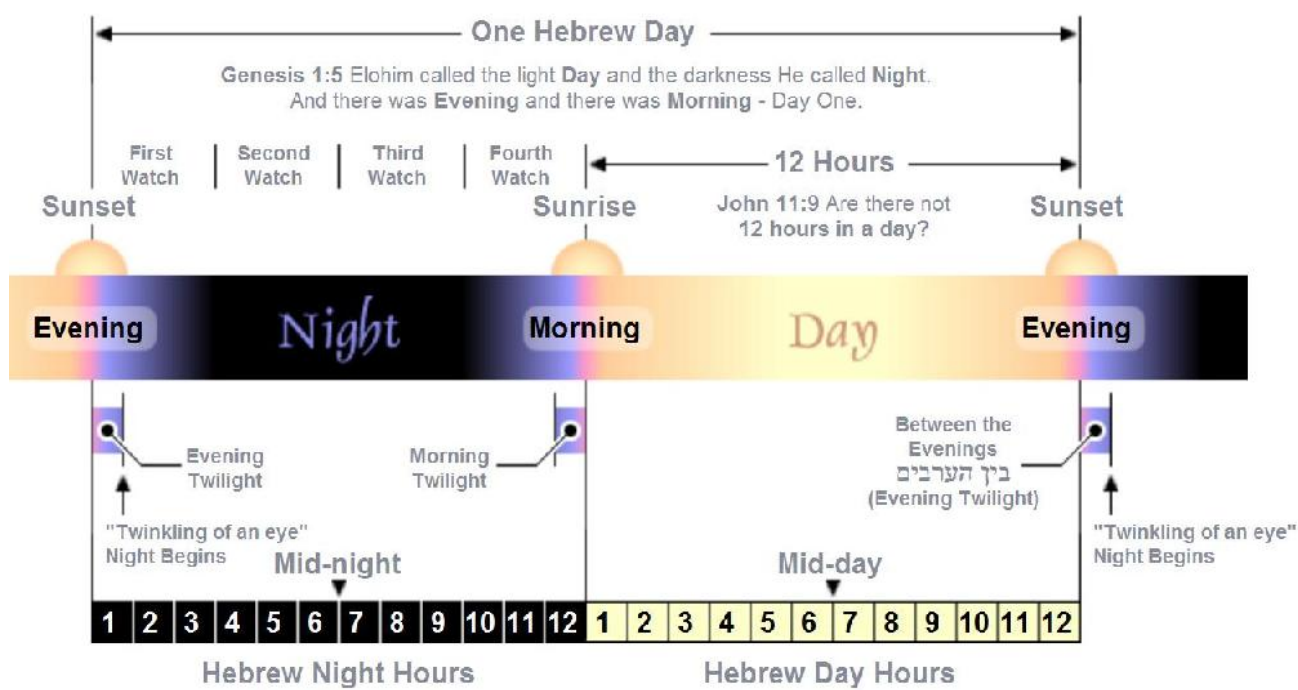

Un giorno per gli ebrei consiste sempre in 12 ore, le ore eguali dei latini. Il punto intermedio delle 12 ore notturne è chiamato mezzanotte, e cade esattamente a metà tra il tramonto e l'alba e separa la sesta dalla settima ora della notte. Analogamente il mezzodì è esattamente a metà tra alba e tramonto, come nel nostro concetto di mezzogiorno locale.

---

[18] Mi fa piacere qui ricordare il prof. Alessandro Cacciani (1938-2007), astronomo a Monte Mario, fisico solare e professore di spettroscopia alla Sapienza, direttore del G28 «Laboratorio di Fisica Solare». Cacciani era un astronomo famoso in tutto il mondo per le sue invenzione del filtro magneto-ottico (MOF, cfr. www.icra.it/solar ), di casa nei maggiori osservatori astronomici solari del Mondo. Si occupò della Sindone, propose un trattamento dell'immagine che dava un effetto tridimensionale che evidenziava il colpo al naso ricevuto dall'Uomo della Sindone. Cacciani nel 1978 ebbe l'occasione insieme ad altri componenti del Centro Romano di Sindonologia di studiare la Sindone da vicino per 3 ore, subito dopo il team dello STURP. Si occupo' anche dell'astronomia della Sindone, come mi è stato riferito dal Padre Gianfranco Berbenni, spiegando proprio il fenomeno della terza stella che sto qui esaminando. Negli ultimi anni della sua vita anche io ho contribuito alla vita del G28 portandovi decine di studenti per il corso di Laboratorio di Astrofisica del prof. De Bernardis a partire dal 2002. Cacciani era molto mattiniero, ed io insegnavo a scuola, quindi in quei cinque anni non abbiamo potuto condividere tutto. Ho avuto l'onore di mostrargli quello che avevo imparato sulla meridiana di Santa Maria degli Angeli proprio nel Luglio del 2006, quando le ultime grandi macchie del ciclo 23 si riuscivano ad osservare al foro stenopeico, senza alcuna lente obiettiva. L'anno successivo un tumore lo ha costretto a dei cicli di chemioterapia, per cui lavorava a casa. Il suo ottimismo non cessava mai, e la sua voce al telefono era sempre positiva. La sua morte è giunta come un fulmine a ciel sereno, e ai suoi funerali alla chiesa di N. Signora del Rosario a Prati ho saputo delle sue attività parallele, anche di Sindonologo.

Un modo per misurare le ore del giorno è usando un orologio solare con dodici divisioni equiangolari.

L'intervallo di tempo tra il tramonto del Sole e l'inizio della notte è chiamato anche «le due sere», a questo si riferiscono i samaritani quando parlano del sacrificio dell'agnello pasquale.[19]

L'inizio della notte, poi, avviene in un battere di ciglio: *the twinkling of an eye*.

Alla domanda su come si debba identificare la terza stella ogni rabbino ha il suo metodo, o si riferisce a quello di qualche altro rabbino. Questa situazione è tipica del mondo ebraico, dove non esiste una guida suprema, ma ciascuna comunità si riferisce al proprio rabbino, e questi casomai si riferisce ad altri rabbini di comunità più grandi. Alcuni rabbini sono più importanti di altri in base alla tradizione ed alla consistenza della propria comunità. Il rabbino capo di Roma, ad esempio, appartiene ad una comunità che esiste da più di 2000 anni, perciò è particolarmente autorevole.

La stessa situazione si trova sul web, andando a cercare la soluzione per la terza stella.

È evidente che i vari rabbini sono coscienti che in presenza di astri molto luminosi come Giove e Venere, magari in Inverno con Sirio già sopra l'orizzonte, la terza stella comparirebbe molto prima che in una sera autunnale senza questi due pianeti sopra l'orizzonte. Allora parlano di stelle di media grandezza, senza però dare indicazioni precise a riguardo, anche perché le stelle visibili cambiano con il cambiare della stagione.

Questa clausola, che siano di media grandezza, fa sì che si debbano escludere le stelle di prima grandezza e i pianeti, che di solito sono anche più luminosi.

---

[19] Dopo le invocazioni preliminari il sacerdote che presiede recita lentamente, con il suo seguito, le parole dell'immolazione: Shem, disse a Mosè ed Aronne nel paese d'Egitto: questo mese sarà per voi il primo dei mesi... dite dunque a tutta l'assemblea dei figli d'Israele... che si procurino ciascuno un capo di piccolo bestiame per famiglia... voi lo sceglierete fra i montoni o le capre. Voi lo custodirete fino al quattordicesimo giorno del mese; allora l'assemblea intera dei figli d'Israele lo sgozzerà entro le due sere (Es 12, 1-6). M. Baillet, *op. cit.* (1986).

Allora quando d'inverno compaiono Sirio e Orione (con ben cinque stelle di prima grandezza), oppure Vega, Arturo, Altair d'estate non è ancora nato il giorno successivo, ma poco dopo quando compaiono le stelle di media grandezza, quando ne compaiono 3, è il nuovo giorno.

Correntemente in astronomia si parla di crepuscoli (twilight), distinguendoli in crepuscolo civile, nautico ed astronomico, a seconda che il Sole sia rispettivamente 6° sotto l'orizzonte, 12° e 18° sotto l'orizzonte.

Nei paesi prossimi ad avere il Sole a mezzanotte, come Helsinki e Stoccolma, nelle notti vicine al solstizio d'estate, il Sole non va mai sotto i 12° e la notte non è mai scura. Ancora più a Nord la terza stella può non comparire per mesi, ed i rabbini dovranno adottare altri accorgimenti.

Per i crepuscoli si possono calcolare agevolmente le effemeridi.

Alla latitudine di Roma si puo' considerare che ogni fase del crepuscolo dura all'incirca mezz'ora, un po' di più d'estate e un po' di meno d'inverno.

Infine credo di dare una testimonianza sensata del concetto di «battere di ciglio», spazio di tempo entro il quale il nuovo giorno entra con l'apparire della terza stella.

Il battito di ciglio è tra la prima e la terza stella di media grandezza che sono apparse. Stavo osservando la pioggia delle Draconidi, stelle cadenti il cui radiante si trova nel Dragone, dal volo Alitalia AZ790 Roma – Pechino, dell'8 ottobre 2011. La notte, a causa del volo e della rotta molto a Nord (fino a sorvolare Krasnojarsk e Irkutsk presso il lago Bajkal), durava quasi 7 ore. Avevo la visuale al Nord, con le costellazioni delle Orse e del Dragone. In particolare stelle di terza grandezza, e qualcuna di seconda (Orsa Maggiore).

La notte a quelle latitudini sarebbe durata quasi 13 ore, essendo ancora vicini all'equinozio d'autunno. Per cui gli intervalli di tempo che ora riporto, dal mio notes vanno moltiplicati per due per avere un valore di riferimento per un osservatore a terra.

| Ora | fenomeno |
|---|---|
| 16:46 | Tramonto (Brašov, Romania) |
| 17:00 | Prima stella (Giove, Capella) |
| 22:55 | Vedo bene la Stella di Gerberto[20] (5.35, quinta grandezza) |
| 23:00 | Distinguo la quarta grandezza |
| 23:04 | Arrivo a vedere la terza grandezza |
| 23:06 | Vedo solo stelle più brillanti della seconda grandezza |
| 23:10 | La magnitudine limite è ancora scesa a 1.5 solo le più brillanti dell'Orsa maggiore |
| 23:39 | Sorgere del Sole |

Dal tramonto all'alba sono passate 6h53m, se fossi stato a terra in una latitudine intermedia tra tutte avrei sperimentato una notte di 12h50m, quindi tutti gli intervalli di tempo misurati, come quello tra le 23:04 e le 23:06 pari a 2 minuti, vanno moltiplicati per 1.865; l'intervallo di tempo tra la sparizione delle stelle di terza grandezza e quelle di seconda osservato a terra sarebbe stato di 3 m 44 s.

Nello spazio di 3 m 44 s si possono vedere tutte le stelle di terza grandezza, che prima non si vedevano per niente: è in questi tre minuti che, in un batter di ciglio, la terza stella compare.

Infine la differenza di tempo tra l'apparizione (sparizione in questo caso) della terza stella ed il sorgere del Sole pari a 34 minuti durante il volo, corrisponde a terra ad 1h 04m. Quel giorno 1 ora e 4 minuti prima dell'alba il Sole, a quella latitudine, era a 9°36' gradi sotto l'orizzonte, una posizione intermedia tra il crepuscolo civile e quello nautico.

Possiamo assumere che questa sia la posizione del Sole perché si possa avvistare la terza stella, con un'incertezza di 15' cioè ¼°, dovuti ai due minuti trascorsi tra le 23:04 e le 23:06 del mio orologio. Quindi Sole a -9°36±15' per l'inizio del riposo sabbatico.

---

[20] Si tratta della stella che individuava il Polo Nord Celeste nel 978, oggetto di una lettera scritta da Gerbert d'Aurillac a Costantino di Fleury. È la stella HR 4893, e nelle osservazioni delle meteore la usavo per tenere sotto controllo la magnitudine limite. Si veda il report sulle Draconidi 2011 su C. Sigismondi, *Airborne observation of 2011 Draconids meteor outburst: the Italian mission* arxiv1112.4873 (2011).

### 5.1 La stella Polare

Il modo più semplice di valutare l'altezza del Sole sotto l'orizzonte al momento della terza stella è aspettare di vedere la Polare dopo il tramonto.

Se consideriamo, infatti, la stella Polare come una stella di media grandezza (la sua magnitudine è 2.12), ho verificato da Roma, il 17 novembre 2012, che è già visibile insieme a Kochab (2.08), la β dell'Orsa Minore alle 17:25, con il Sole già sotto di 7°19'. Entrambe queste stelle sono di seconda grandezza.

Il cielo a occidente era ancora chiaro, poiché era appena passato il crepuscolo civile, che, va ricordato, essere il momento di accendere le luci delle auto per il codice della strada (prima che la guida diventasse un'attività da film poliziesco, per cui le luci devono essere sempre accese anche di giorno nelle strade principali).

Il vantaggio di usare la stella Polare è che questa si trova sempre nello stesso punto, ed è facile aspettarla mentre fa buio.

### 5.2 Quando già splendevano le luci del Sabato

Si noti *Lc* 23:54: "Era il giorno della Preparazione, e stava per cominciare il sabato" (*NR*). *CEI* traduce così: "Era il giorno della parascève e *già splendevano le luci del sabato*". Qualcuno tenta di tradurre "cominciava ad *albeggiare* il sabato". Il testo greco originale ha: ἐπέφωσκεν (*epèfosken*). Si tratta del verbo greco ἐπιφώσκω (*epifòsko*) che indica il "crescere di luminosità", che è una forma di ἐπιφαύσκω (*epifàusko*), "spendere sopra", a sua volta una forma di ἐπιφαίνω (*epifàino*) che indica l'"apparire", anche di stelle (*Vocabolario del Nuovo Testamento*).

Nelle Scritture greche il verbo ἐπιφώσκω (*epifòsko*) lo troviamo solo due volte: qui in *Lc* 23:54 e in *Mt* 28:1, troppo poco per desumerne il pieno significato. Per comprenderne appieno il significato dobbiamo quindi ricorrere al vocabolario di greco. Il *Rocci*, il più autorevole vocabolario greco italiano, dà due definizioni: a) cominciare a splendere, b) far splendere.

Il significato di "cominciare a splendere" è esattamente quello dei due passi biblici che contengono il verbo. In *Mt* 28:1 si ha: "Dopo il sabato, quando *cominciava a sorgere la luce* [τῇἐπιφωσκούσῃ (*te apifoskùse*); letteralmente: "alla cominciante a splendere"] del primo giorno della settimana". In *Lc* 23:54 si ha lo stesso significato: "Giorno era di preparazione e sabato *cominciava a splendere* [ἐπέφωσκεν (*epèfosken*)]" (*Nuovo Testamento Interlineare*, San Paolo). Che cosa "cominciava a splendere" in quel giorno della Preparazione delle Pasqua? *TNM* rende così: "Ora era il giorno della Preparazione e *si appressava la luce serale del sabato*". Iniziavano ad accendersi le luci per il giorno festivo di Pasqua (15 *nissàn*), in cui sarebbe poi stato vietato accendere fuochi (*Es* 35:3). Era verso sera. Infatti, "le donne, che erano venute con lui dalla Galilea", ebbero il tempo di seguire Giuseppe d'Arimatea (*Lc* 23:50-53; *Mt* 27:57-60; *Mr* 15:43-46; *Gv* 19:38-42) e di guardare "la tomba commemorativa e come era posto il suo corpo[di Yeshùa]"; poi "tornarono a preparare aromi e oli profumati. Ma il sabato, naturalmente, si riposarono secondo il comandamento". – *Lc* 23:55,56.

*Mt* 27:57 specifica che era "tardo pomeriggio", anzi, il testo greco dice: Ὀψίας δὲ γενομένης (*Opsìas de ghenomènes*), "sera poi fattasi". Si parla qui di Giuseppe d'Arimatea che "andò da Pilato e chiese il corpo di Gesù" (v. 58). Il passo di *Lc* 23:54 dimostra quindi che il giorno finisce di sera, quando ne inizia uno nuovo.[21]

---

[21] Il testo del par. 5.1 è tutto ripreso dal sito http://www.biblistica.it/1/il_giorno_biblico_e_il_suo_inizio_5167654.html creato dal dr. Gianni Montefameglio. L' autore propende per una cronologia diversa: morte di Gesù il mercoledì 14 nisan, resta nel sepolcro 3 giorni e 3 notti come Giona, 14-15;15-16 e 16-17 e risorge all'alba del 18. Mercoledì, 14 nissàn. Prima che faccia buio Yeshùa è posto nel sepolcro. Giovedì, 15 nissàn. "Sabato" (giorno festivo). Yeshùa è nel sepolcro: al tramonto si compie il primo giorno e la prima notte. Venerdì, 16 nissàn. "Passato il sabato, Maria Maddalena e Maria Giacomo e Salome comprarono aromi" (Mc 16:1). "Poi tornate, prepararono aromi e profumi". – Lc 23:56. Yeshùa è nel sepolcro: al tramonto si compie il secondo giorno e la seconda notte. Sabato, 17 nissàn. Sabato settimanale. "Il sabato si riposarono, secondo il comandamento". – Lc 23:56. Yeshùa è nel sepolcro: al tramonto si compie il terzo giorno e la terza notte. Yeshùa viene resuscitato. Domenica, 18 nissàn. Primo giorno della settimana (nostra domenica): le donne trovano la tomba vuota.

L'ipotesi è suggestiva per il compimento dei 3 giorni nel sepolcro come Giona e per l'attività delle donne con gli aromi, ma manda a pallino tutta la cronologia, poiché il mercoledì 14 nissàn arriva solo negli anni 34 o 35, a seconda dell'algoritmo usato per stabilire, a posteriori, l' avvistamento della prima falce di Luna crescente.

A queste datazioni giunse Isaac Newton che studiò approfonditamente l'argomento nel 1733.

# 6. Il Sole e l'Eclissi del Venerdì Santo nelle catechesi di san Cirillo di Gerusalemme

Chi innalza lo sguardo verso il Sole può non ammirarne il modo con cui è regolato?
Al suo primo apparire infatti può sembrare una cosa di poco conto, mentre di fatto ha una potenza veramente grande.
Esso ci dà luce da quando spunta in Oriente fino a che tramonta in Occidente.

Quando il Salmista dice che esso sorge al mattino «come sposo dal talamo», dà un'immagine dei suoi raggi temperati che non offendono gli occhi degli uomini al suo sorgere, perché al suo apparire «come uno sposo» lo troviamo piacevole; solo quando dirige i suoi cavalli verso mezzogiorno per lo più ci ripariamo dai suoi raggi infuocati.

Osserva soprattutto il modo in cui esso si regola, benché non sia il Sole ad assegnare la regola ma Colui che gli ha definito il corso da seguire.

D'estate, innalzandosi, prolunga le giornate per dare agli uomini più tempo per lavorare; d'inverno invece accorcia il suo corso non certo per allungare il tempo del freddo, ma perché le notti facendosi più lunghe cooperino con gli uomini favorendone il riposo e con la terra favorendone la fruttificazione.

Osserva anche l'armonia con cui i giorni si equilibrano tra di loro: d'inverno si accorciano e d'estate si allungano, in primavera e in autunno si gratificano l'un l'altro perseguendo parità di lunghezza. Alla stessa maniera si regolano le notti. Lo dice il Salmista: «Il giorno al giorno affida il messaggio, e la notte alla notte trasmette notizia».

Giorni e notti con il loro ordinato procedere all'unisono gridano agli eretici che non vi è altro Dio al di fuori del Creatore di tutte le cose che ad esse ha dato regola e ordine.
Ma gli eretici non hanno orecchie.   [Cat. 9, 6]

In questo passo si vede chiaramente che Cirillo conosce bene la descrizione delle stagioni attraverso i percorsi celesti del Sole, ed è implicito nel suo sostrato culturale anche il concetto delle ore disuguali legato alla diversa durata del dì d'estate e d'inverno.

Così Cirillo argomenta dell'ora sesta in cui con Cristo il sole si eclissò:

Vuoi sapere a che ora esatta il sole fu in eclissi? Alla quinta, all'ottava o alla decima? Dillo con esattezza ai giudei increduli, o profeta.

Il profeta Amos precisa con esattezza quando il sole entrò in eclissi: In quel giorno – oracolo del Signore Dio – farò tramontare il sole a mezzodì e oscurerò la terra in pieno giorno (Am. 8, 9). È quello che dice il Vangelo: all'ora sesta si fece buio.

Quali i comportamenti, e in quale giorno si verificheranno, o profeta? Lo dice in seguito: Cambierò le vostre feste in lutto (Am. 8, 9). Si era allora infatti nella festa di Pasqua e la si celebrava con la prassi degli azzimi (cf. Mt. 27, 2; Mc. 14, 1); perciò il profeta aveva soggiunto: Muterò la festa in un lutto come per il figlio unico, e farò del periodo seguente come un giorno d'amarezza (Am. 8, 10). Nel periodo festivo degli azzimi, di fatto, le loro donne si percuotevano e facevano compianti (cf. Lc. 23, 27), come fecero gli apostoli che di nascosto facevano lamenti dando mirabilmente compimento alla profezia. [Cat. 13, 25]

Perciò, quando venne meno il Sole di giustizia (cf. Mal 4, 2), il sole si eclissò (cf. Lc 23, 45); quando si infranse la Roccia spirituale (cf. 1Cor 10, 4), le rocce materiali si spaccarono (cf. Mt 27, 51); quando il Libero tra i morti (cf. Sal 88 / 87, 6; Zac. 9, 11; Is. 53, 4-5) andò a liberare da una fossa senz'acqua i prigionieri che gli appartenevano, i sepolcri si aprirono (cf. Mt 27, 32). [Cat. 13, 34]

In entrambi i passi della Catechesi tredicesima, sulla Croce, Cirillo parla di eclissi. Per lui il problema astronomico di un'eclissi solare in tempo di Luna piena non sussiste. E torna a parlare di eclissi nella menzione della retrocessione del Sole sulla meridiana di 10° come segno concesso ad Ezechia.

Il sole fece una retrocessione per Ezechia, e invece si eclissò per il Cristo; in entrambi i fenomeni di retrocessione e di eclissi, diede un segno, benché di diverso significato nei riguardi di Ezechia e di Gesù: per Ezechia fu segno della sentenza divina revocata, per Gesù fu forse segno della remissione avvenuta dei peccati. [Cat. 1, 15]

Queste catechesi sono state proclamate proprio sulla roccia del Golgota, da 13 anni riaperta al culto cristiano, roccia testimone, come il Sole del Venerdì Santo, della Passione, Morte e Risurrezione del Signore:

Qui fu crocifisso perché noi fossimo liberati dei nostri peccati, non certo per i suoi; qui dopo essere stato dagli uomini disprezzato e schiaffeggiato come un semplice uomo, fu riconosciuto dal creato come Dio, quando il sole vedendo il suo Signore vilipeso vacillò e non soffrendo più quella vista abbandonò il suo posto.[Cat. 4, 10]

La stessa roccia dove il corpo morto di Cristo

Come uomo fu veramente deposto in un sepolcro di pietra: per questo ogni roccia tremò e si spaccò. [Cat. 4, 10]

e che ci fa proclamare che è

Vera la morte di Cristo,
vera la separazione della sua anima dal suo corpo,
vera anche la sepoltura del suo santo corpo
avvolto in una sindone (cf Mt 27, 59).
In lui tutto è veramente avvenuto,
per voi invece non è avvenuta
che una somiglianza della sua morte e della sua passione. [Cat. 20, 7]

## 7. L'eclissi di Sole secondo Luca

Καὶ ἦν ἤδη ὡσεὶ ὥρα ἕκτη, καὶ σκότος ἐγένετο
ἐφ' ὅλην τὴν γῆν ἕως ὥρας ἐνάτης |
τοῦ ἡλίου ἐκλιπόντος
(Lc 23, 44-45a).

A mio avviso[22] la traduzione deve rispettare l'ordine con cui si presentano nel greco i vari termini: "Era circa mezzogiorno e si fece buio su tutta la terra fino alle tre del pomeriggio, quando il Sole si eclissò". In questo modo il genitivo assoluto viene rispettato sia nella sua funzione che nella sua posizione nell'economia del racconto.

Il verbo εκλειπω associato ad ηελιος indica proprio l'eclissarsi del Sole[23]. Si noti che dalla radice di εκλειπω proviene anche il termine "eclittica", che nella sfera celeste è il cerchio massimo che il Sole percorre nel suo cammino apparente nel corso dell'anno e che prende questo nome proprio dal fatto che quando la Luna attraversa l'eclittica, nei nodi della sua orbita, e contemporaneamente si trova in congiunzione o in opposizione al Sole avviene un'eclissi.

---

[22] C. Sigismondi, *Tematiche astronomiche nei Vangeli in prospettiva cristologica*, Pontificia Università Lateranense (1998).
[23] Cfr. Tucidide 2, 28, 1 (dopo mezzogiorno, nella *Guerra del Peloponneso*); 7, 50, 4; Platone, *Fedone*, 99d; Plutarco, *Aless.*, 31, 8.

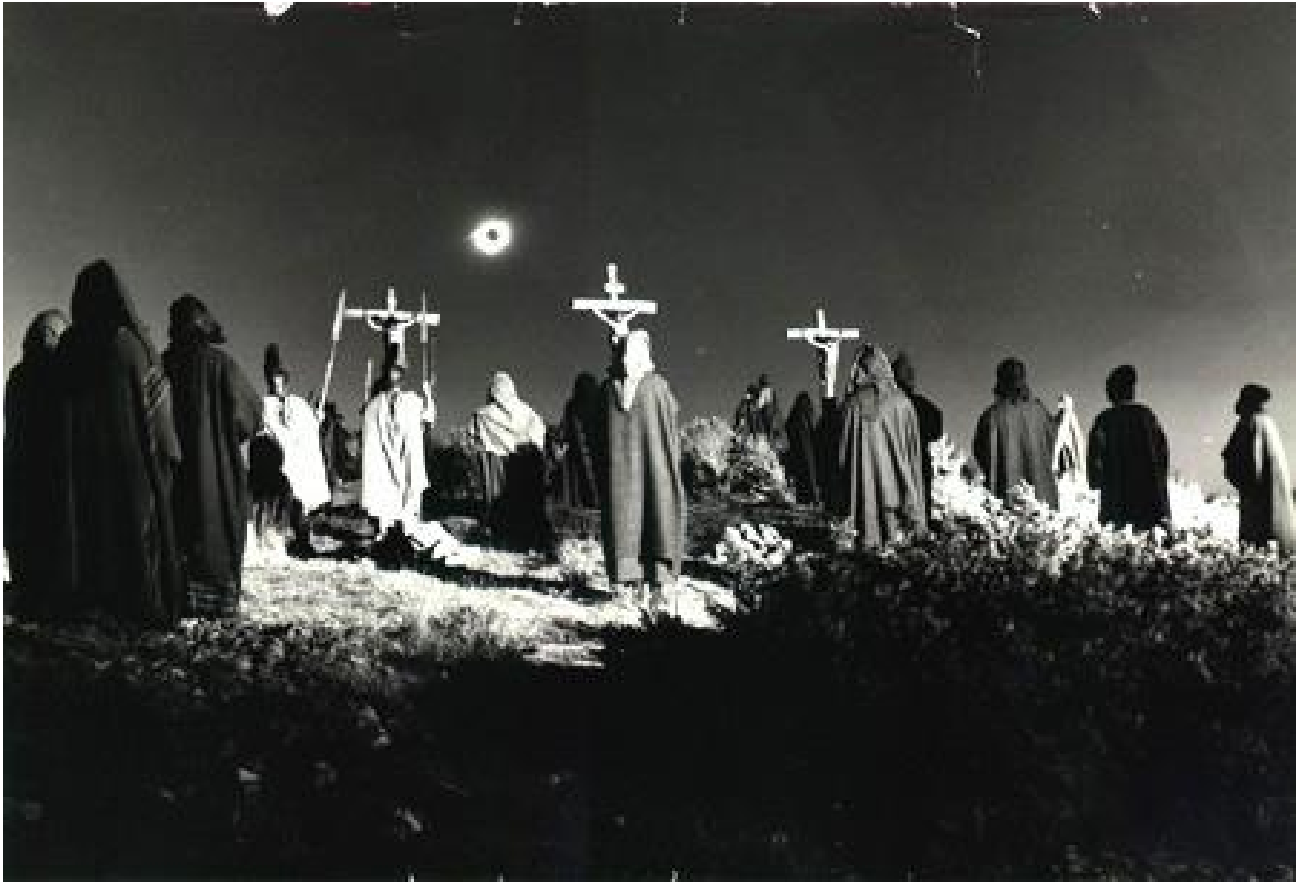

Sul set del film Barabba, di Dino de Laurentiis, il fotografo Augusto di Giovanni ha catturato l'eclissi del 15 febbraio 1961, in Toscana, mentre veniva girata la scena della Crocifissione.[24]

---

[24] http://www.andreaplanet.com/italy/

Questo è stato il fenomeno descritto da Luca? Un'eclissi, nella sua fase totale, non dura che pochi minuti. La più lunga teoricamente possibile è di 7 m 40 s. Nel 1991 in Messico si verificò la più lunga degli ultimi secoli di 7m 11s. Niente a che vedere con le tre ore di oscurità durante la crocifissione, dall'ora sesta all'ora nona, descritte dagli evangelisti. L'ipotesi scientifica sull'origine dell'oscurità è di origine meteorologica, non potendosi verificare un'eclissi di Sole con la Luna piena. L'altra ipotesi è il *topos letterario*, come quello delle «portentose e lunghe eclissi di Sole»[25] verificatesi alla morte di Cesare, ovviamente impossibili visto che la probabilità media di vedere un'eclisse totale da uno stesso luogo è di una ogni 400 anni, e per Roma attendiamo dal 1567 la prossima eclissi totale nel 2187!

Già lo storico Giulio Africano mostrò il problema astronomico, e Origene addirittura sosteneva che la menzione dell'eclissi solare è stata frutto dell'azione di detrattori per portare la contraddizione nella verità evangelica.

Alcuni a partire da questo testo [Mt 27, 45] calunniano la verità evangelica dicendo: come può essere vero, secondo il teso, che *sono scese le tenebre sopra tutta la terra dall'ora sesta fino all'ora nona*, fatto che non la storia non riporta. D'altronde l'eclissi di Sole è sempre accaduta, da che mondo è mondo, al suo tempo [...], e non in altro tempo se non quello della congiunzione del Sole e della Luna [...]. Nel tempo della passione di Cristo, è evidente che era tempo pasquale [...] quando la Luna è piena ed è visibile per tutta la notte. In quale modo dunque sarebbe potuta accadere un'eclissi di Sole con la Luna piena?[26]

Luca, medico, scrupoloso redattore del terzo Vangelo e degli Atti degli Apostoli, ha effettivamente e, con tutta probabilità, volutamente scritto un resoconto di eclissi culminata alle tre del pomeriggio con la totalità.

Ma perché? La mia ipotesi resta quella che egli stesso potrebbe essere stato testimone di un'eclissi totale visibile dal suo paese di origine, Antiochia sull'Oronte, capitale della provincia romana della Siria e terza città dell'impero romano dopo Roma ed Alessandria, accaduta il 24 novembre del 29 d. C. alle 11 e 15 del mattino, quando la totalità è durata al massimo 1 minuto e 1/2 per gli osservatori posti esattamente al centro della fascia di totalità. Quest'eclissi fu l'unica del I secolo d. C. visibile in quell'area[27]; oppure Luca può esserne stato informato da altre persone. E le sensazioni provate durante l'eclissi totale al momento del "Sole nero" sono state volutamente associate alla descrizione del momento della morte del Signore.

---

[25] Plinio il Vecchio, *Naturalis Historia* 2.30, 97.
[26] Origene, *Commentariorum series* 134; *in Matt.* 27, 45 [*Die Griechischen Christlichen Schriftsteller* (Berlin) 38.271-74].

Luca ha voluto dare una connotazione particolare a quella tenebra del Venerdì Santo, la più sconvolgente possibile, basandosi forse su una sua esperienza diretta, lui che non era stato testimone della Crocifissione.

Un'ultima considerazione riguarda ancora Luca, come pittore. Nelle icone bizantine la gloria divina talvolta è rappresentata sotto forma di mandorla dorata, e in alcuni casi nera. Il nero è indice dell'essere completamente abbagliato come quando si guarda il Sole ed esso satura completamente i nostri foto-recettori. Durante l'eclissi è proprio l'unico momento in cui si può fissare il Sole senza esserne abbagliati, circondato da una mandorla che è la corona solare.

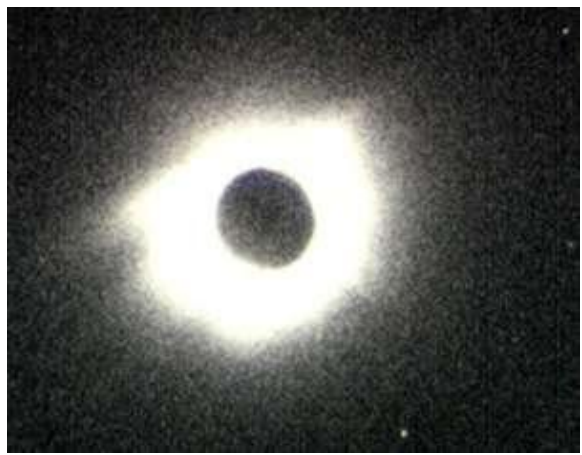

L'immagine originale della crocifissione cinematografica del 1961, presa con una Hasselblad mostra anche qualche dettaglio della corona solare ed alcuni astri, visibile in questo ingrandimento.

---

[27] Sawyer, J. F.A., *Why a solar eclipse mentioned in the passion narrative (Luke XXII. 44-5)?*, Journal of Theological Studies New series **23** 124-128 (1972), affermazione poi verificata da me col software Occult 4.

**Bibliografia**

**Siti web**